\documentclass{aa}
\usepackage[varg]{txfonts}
\usepackage{amsmath}
\usepackage{amsfonts}
\usepackage{amssymb}
\usepackage{upgreek}
\usepackage{natbib}
\usepackage{xcolor}
\usepackage{hyperref}
\usepackage{graphicx}
\hypersetup{
    colorlinks=true,
    linkcolor=blue,
    citecolor=blue,
    filecolor=magenta,      
    urlcolor=cyan,
}

\bibpunct{(}{)}{;}{a}{}{,}

\begin{document}

\title{Transition to turbulence in nonuniform coronal loops driven by torsional Alfv\'{e}n waves.}
\subtitle{II. Extended analysis and effect of magnetic twist}

\author{Sergio D\'{i}az-Su\'{a}rez \inst{\ref{inst1},\ref{inst2}} \and Roberto Soler \inst{\ref{inst1},\ref{inst2}}}
\institute{Departament de F\'{i}sica, Universitat de les Illes Balears, E-07122, Palma de Mallorca, Spain \label{inst1} \and Institute of Applied Computing \& Community Code (IAC3), Universitat de les Illes Balears, E-07122, Palma de Mallorca, Spain \label{inst2}; \email{s.diaz@uib.es}}

\date{Received 02 June 2022 /Accepted 08 July 2022}
\abstract{It has been shown in a previous work that torsional Alfv\'{e}n waves can drive turbulence in nonuniform coronal loops with a purely axial magnetic field. Here we explore the role of the magnetic twist. We modeled a coronal loop as a  transversely nonuniform straight flux tube, anchored in the photosphere, and embedded in a uniform coronal environment. We considered that the magnetic field is twisted and control the strength of magnetic twist by a free parameter of the model. We excited the longitudinally fundamental mode of standing torsional Alfv\'{e}n waves, whose temporal evolution was obtained by means of high-resolution three-dimensional ideal magnetohydrodynamic numerical simulations. We find that phase mixing of torsional Alfv\'{e}n waves creates velocity shear in the direction perpendicular to the magnetic field lines. The velocity shear eventually triggers the Kelvin-Helmholtz instability (KHi). In weakly  twisted magnetic tubes, the KHi is able to grow nonlinearly, and subsequently, turbulence is driven in the coronal loop in a similar manner as in the untwisted case. When the magnetic twist remains weak, it delays the onset of the KHi and slows the development of turbulence down. In contrast, magnetic tension can suppress the  nonlinear growth of the KHi when the magnetic twist is strong enough, even when the KHi has locally been  excited  by the phase-mixing shear. Thus, turbulence is not generated in strongly twisted loops.}


\titlerunning{Torsional Alfv{\'e}n waves in twisted magnetic fields}
\authorrunning{S. D\'{i}az-Su\'{a}rez \& R. Soler}
\maketitle

\defcitealias{Diazsoler21b}{I}

\section{Introduction}
High-resolution and high-cadence  observations have shown the ubiquity of   magnetohydrodynamic (MHD) waves throughout the solar atmosphere \citep[see, e.g.,][]{Aschwanden99,Nakariakov99,Tomczyk07,Depontieu07,Mcintosh11,Morton15,Jafarzadeh17,Srivastava17}. The dissipation of MHD waves could play an important role in the heating of the solar corona \citep[see, e.g.,][]{Hollweg78,Cranmer05,Cargill11,Mathioudakis13,Soler19,Tom20,Nakariakov20}, and the acceleration of the solar wind \citep[see, e.g.,][]{Charbonneau95,Cranmer09,Matsumoto12,Shoda18}.

Torsional Alfv\'{e}n waves are a subtype of axisymmetric Alfv\'{e}n waves in cylindrical flux tubes. They are nearly incompressible, the restoring force is the magnetic tension, and their direction of polarization is perpendicular to the magnetic field lines. This type of waves  produces  axisymmetric perturbations in the perpendicular components of the velocity and magnetic field. Furthermore, unlike Alfv\'{e}n waves with other azimuthal symmetries,  torsional Alfv\'{e}n waves do not couple with magnetoacoustic waves when the magnetic field is straight \citep{Goossens11}. Torsional Alfv\'{e}n waves have been reported in bright points \citep{Jess09} and in a coronal active region structure \citep{Kohutova20}. The torsional motions found in the chromosphere and transition region by \citet{DePontieu12,DePontieu14} can also be interpreted as torsional Alfv\'en waves.  \citet{Aschwanden20} interpreted  oscillations in the magnetic energy during solar flares 
as torsional Alfv\' {e}n waves. This type of waves has also been reported in the solar wind during the interaction between coronal mass ejections \citep{Raghav18}.

In coronal loops, which are closed magnetic structures with  two footpoints anchored in the solar photosphere, there is no report of such waves. Nonetheless, torsional Alfv\'{e}n waves can be excited at their footpoints through vortex or twisting plasma motions in the solar photosphere \citep[see, e.g.,][]{Fedun11,Shelyag11,Shelyag12,Wedmeyer12,Mumford15,Srivastava17b}. Recently, \citet{Soler21} have investigated the propagation of torsional Alfv\'{e}n waves from the photosphere to a coronal loop. As the waves reach the coronal loop, they can resonate with standing modes of the loop and drive global torsional oscillations \citep[see also][]{Hollweg84,Hollweg84b}. \citet{Soler21} found that a large amount of energy can be transmitted  at coronal heights despite the chromospheric filtering. Particularly, they reported that the energy flux is channeled mostly through the fundamental standing torsional mode of the loop.

Inspired by these results, in \citeauthor{Diazsoler21b} (\citeyear{Diazsoler21b}; hereafter Paper \citetalias{Diazsoler21b}), we investigated the nonlinear evolution of torsional Alfv\'{e}n waves in a straight coronal flux tube with a constant axial magnetic field. In Paper \citetalias{Diazsoler21b} (see also the torsional model in \citet{Guo19}), we showed that the phase mixing of the torsional Alfv\'{e}n waves generates azimuthal shear flows. These flows eventually excite the Kelvin-Helmholtz instability (KHi), as \citet{HeyvaertPriest83} and \citet{Browning84} predicted analytically. Phase mixing is a linear phenomenon that occurs when waves in adjacent radial positions oscillate with different frequencies. The cause of phase mixing is a frequency continuum, whose origin lies in the inhomogeneities in density and/or magnetic field \citep[see, e.g.,][]{HeyvaertPriest83,Nocera84,Moortel00,Smith07,Prokopyszyn2019}. Phase mixing increases the values of vorticity and current density and generates small scales, although at a relatively slow pace.

The KHi triggered by phase mixing can evolve nonlinearly by forming large eddies that break into smaller eddies. This process initiates and drives turbulence. Turbulence accelerates the energy cascade to small scales, which might enhance the efficiency of wave energy dissipation \citep{Hillier20}. There is evidence that coronal loops may be in a turbulent state \citep{Demoortel14,Hahn14,Liu14,Xie17}. The generation of turbulence is also a result obtained by numerous studies of numerical simulations of kink oscillations of coronal loops \citep[see, e.g.,]     []{Terradas08,Terradas18,Antolin15,Magyar16,Howson17a,Karampelas19,Antolin19}.

We here extend the investigation of Paper~\citetalias{Diazsoler21b} by replacing the constant axial magnetic field by a twisted magnetic field. The existence of twisted coronal loops has been confirmed by observations \citep[see, e.g.,][]{Kwon08,Aschwanden12,Thalmann14,Aschwanden19}. The behavior of linear MHD waves in twisted flux tubes has extensively been  investigated analytically or semianalytically \citep[see, e.g.,]               []{Bennet99,Erdelyi06,Erdelyi07,Erdelyi10,Rudermann07,Karami09,Karami10,Terradas12,Ebrahimi17,Ebrahimi22}. The nonlinear evolution of kink MHD waves in twisted flux tubes has  also been investigated numerically \citep{Howson17a,Terradas18}. Here, we explore how  magnetic twist affects the nonlinear evolution of torsional Alfv\'en waves.

\section{Numerical setup}
\label{Sect_Num}
\subsection{Initial configuration} \label{Subsect_model}

As in Paper~\citetalias{Diazsoler21b}, we use the standard coronal loop model \citep[see, e.g.,][]{Edwinroberts83}. The model is made of an overdense cylindrical flux tube of radius, $ R $, and length, $ L, $ embedded in a low-$\beta$ uniform coronal environment. We set $L/R = 10$. We used a shorter loop length than what is reported from observations, typically $L/R \sim 100$. The main reason for this is that we wish to speed up the simulation times. The periods of the standing torsional oscillations are proportional to the loop length. Thus, if a longer loop were considered, the periods and simulation times would be longer, but the dynamics would be essentially the same. The effect of considering larger loop lengths was explored in Paper~\citetalias{Diazsoler21b}.

The loop footpoints were fixed at two rigid walls representing the solar photosphere.  We neglected the curvature of the coronal loop and the thin chromospheric layer at the loop feet.  In Paper~\citetalias{Diazsoler21b} we assumed a uniform magnetic field along the cylinder axis. Here, we improved the model by considering a twisted magnetic field, which is a more realistic representation of the magnetic field in coronal loops, namely
\begin{equation}
    {\bf B} = B_{\varphi}(r) \hat{e}_\varphi + B_{z}(r) \hat{e}_z, 
\end{equation}
where  $B_{\varphi}$ and $B_{z}$ are the  azimuthal and longitudinal components of the background magnetic field in a cylindrical coordinate system denoted by $r$, $\varphi$, and $z$. In our model, the $z$-axis coincides with the flux tube axis. We considered the same force-free twist model as was used in \citet{Terradas18}, namely
\begin{eqnarray}
B_{\varphi}(r)&=&\left\lbrace \begin{matrix} cr/R, & \mbox{if}\;r < R,\label{bphi} \\
 cR/r, & \mbox{if}\;r\geq R,\end{matrix}\right. \\
B_{z}(r)&=&\left\lbrace \begin{matrix} \sqrt{B^{2}_{0}+2c^{2}\left(1-r^{2}/R^{2}\right)}, & \mbox{if}\;r < R, \label{bz}\\
 B_{0}, & \mbox{if}\;r\geq R,\end{matrix}\right.  
\end{eqnarray}
where  $ B_{0} $ is a constant that corresponds to the magnetic field strength at the tube axis, and $c$ is a dimensionless parameter that controls the amount of magnetic twist. 

The highest value of the magnetic twist parameter used here is $ c = 0.4$, which was chosen according to observational constraints. \citet{Aschwanden13p3} computed the twist angle, $ \theta $, in coronal loops as $ \theta=\arctan \sqrt{q_{\mathrm{free}}}$, where $ q_{\mathrm{free}}$ is the free energy ratio. The calculation of $q_{\mathrm{free}}$ can be achieved from observations through line-of-sight magnetograms, the three-dimensional (3D) position of the coronal loop, and a nonlinear force-free field code. Tables 3 in \citet{Aschwanden14} and \citet{Aschwanden16} show that the free energy ratio ranges from 0 to $\sim 0.25 $. Therefore, twist angles smaller than  $ \sim 25^{\circ}$ are expected. The twist angle of our model can be  calculated as
\begin{equation}
\theta=\arcsin\left(\frac{B_{\varphi}(r)}{B(r)}\right),
\label{angles}
\end{equation}
where $ B(r)=\sqrt{B^{2}_{\varphi}\;(r)+B^{2}_{z}\;(r)} $ is the modulus of the magnetic field.
For $c=0.4$, the twist angle is $ \theta =21.8^{\circ}$ at $r=R$, which is slightly smaller than the largest angle inferred by  \citet{Aschwanden14} and \citet{Aschwanden16}. On the other hand, the number of turns of the magnetic field lines over the cylinder length can be computed as
\begin{equation}
N\left(r\right)=\frac{1}{2\pi} \frac{L}{r} \frac{B_{\varphi}\left(r\right)}{B_{z}\left(r\right)},
\label{ntw}
\end{equation}
where the factor $L B_{\varphi}\left(r\right)/rB_{z}\left(r\right) \equiv  \Phi(r) $ is the absolute twist. \citet{Kwon08} considered a sample of 14 loops from  Transition Region And Coronal Explorer (TRACE; \citeauthor{TRACE} \citeyear{TRACE}) observations in the extreme ultraviolet and determined that their absolute twist values ranged  from $0.22\pi$ to $1.73\pi$. This result implies that $0.11 < N < 0.865$.  At $ r=R $ in our model, where $B_{\varphi} $ is maximum, we obtain $ N= 0.64$ for $c=0.4$ and $ L/R= 10$, so that the maximum twist considered here is below the upper limit of the  interval inferred by \citet{Kwon08}.

A twisted magnetic field is prone to develop a kink instability. In the case of uniform twist,  \citet{HoodPriest79} found that the kink instability can develop when the absolute twist is larger than $ 3.3\pi $. With an improved analysis, \citet{HoodPriest81} reduced the threshold value to $ 2.49\pi $. Using the definition of $N$ in Eq.~(\ref{ntw}), we find that the kink instability may appear in our model if $ N \geq 1.65$ according to the critical twist of  \citet{HoodPriest79} and if $ N \geq 1.245 $ according to the critical twist of  \citet{HoodPriest81}. Clearly, we are sufficiently far away from the critical twist for the kink instability because our maximum twist ($ N= 0.64$) is still relatively weak.

We note that we used a preexisting twisted magnetic field in our model. A different way to proceed would be to create twist in the model by dynamically rotating the magnetic field lines anchored in the photosphere, as done in \cite{Ofman09}. This alternative approach imposes a more complex scenario and is not considered here for simplicity.

\begin{figure}[htbp!]
\resizebox{\hsize}{!}{\includegraphics{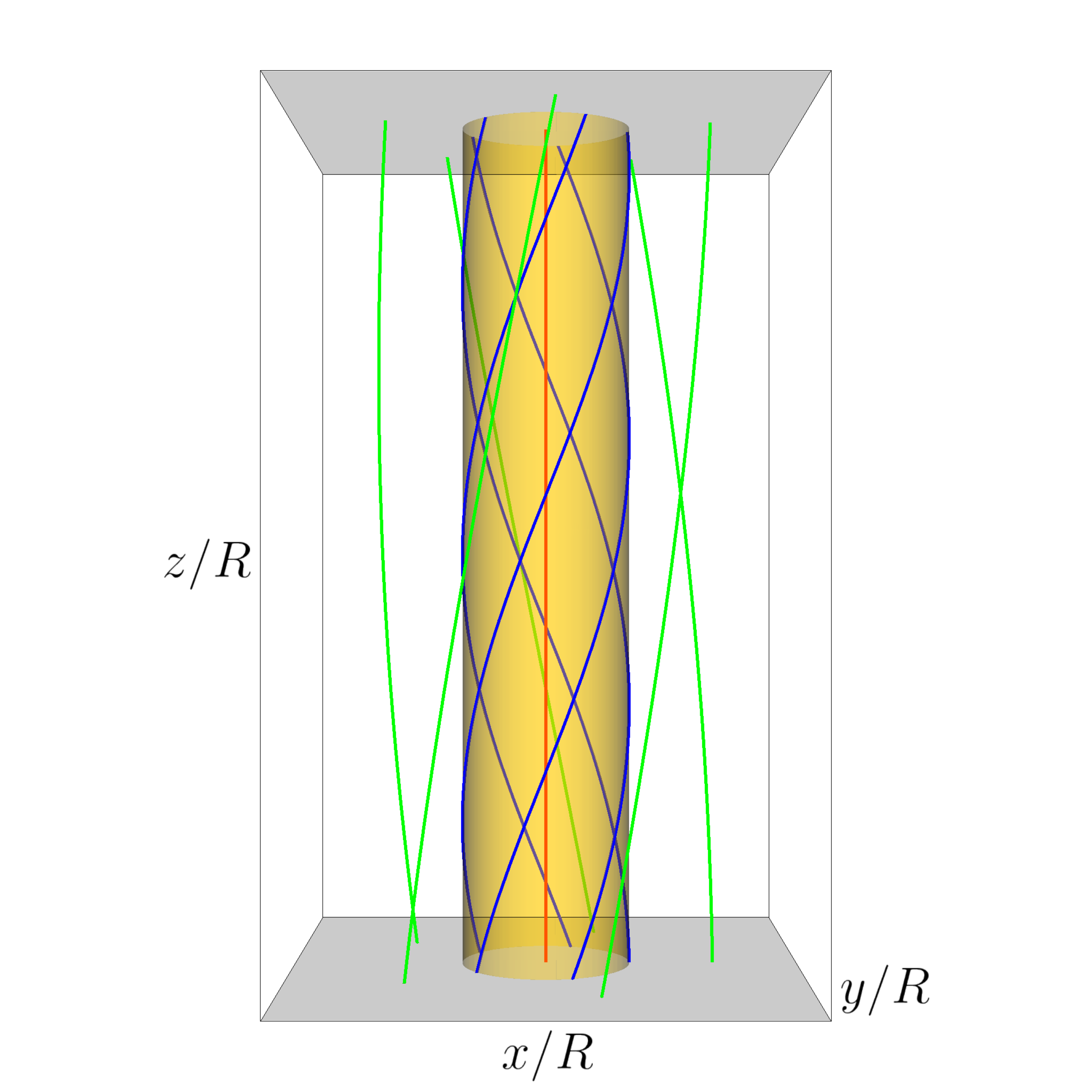}} 
\centering
\caption{Sketch of the coronal flux-tube model. The isosurface of density with value equal to $(\rho_{\mathrm{i}} + \rho_{\mathrm{e}})/2$ is shown in orange. Magnetic field lines located at $r=2R$, $r=R$ and $r=0$ are drawn in green, blue, and red, respectively, for the model with $c=0.4$. The bottom and top gray planes represent the solar photosphere. }
\label{Fig_sketch}
\end{figure}

Since the magnetic field is force free, we considered in the model a background uniform gas pressure, $p_0$. In turn, the equilibrium density, $\rho_0$, was the same as in Paper \citetalias{Diazsoler21b}. Thus, we used
\begin{equation}
\rho_{0}(r)=\left\{
\begin{array}{lll}
\rho_{\mathrm{i}}, & \mbox{if} & r \leq R-\frac{l}{2},\\
\rho_{\mathrm{tr}}(r), & \mbox{if} & R-\frac{l}{2}<r< R+\frac{l}{2}, \\
\rho_{\mathrm{e}},  & \mbox{if} & r \geq R+\frac{l}{2}.
\end{array}
\right.
\label{rhogen}
\end{equation}
In Eq.~(\ref{rhogen}), $ \rho_{\rm i} $ is the density in the uniform inner core, $ \rho_{\rm e} $ is the uniform external density, and $ \rho_{\rm tr}(r) $ is the density in a nonuniform transitional layer of thickness $ l $ that continuously connects the two uniform regions as
\begin{equation}
\rho_{\mathrm{tr}}\left(r\right)=\frac{\rho_{\mathrm{i}}}{2}\left\lbrace \left[1+\frac{\rho_{\mathrm{e}}}{\rho_{\mathrm{i}}}\right]- \left[1-\frac{\rho_{\mathrm{e}}}{\rho_{\mathrm{i}}}\right]\sin\left[\frac{\pi}{l}\left(r-R\right)\right]\right\rbrace,
\label{rhotr}
\end{equation}
where $ l $ may range from 0 (abrupt transition) to $2R$ (fully nonuniform loop). We set $\rho_{\rm i}/\rho_{\rm e}=2$ and $ l=0.4R $. A scheme of our model is depicted in Fig. \ref{Fig_sketch}.

\begin{figure}[htbp!]
\resizebox{\hsize}{!}{\includegraphics{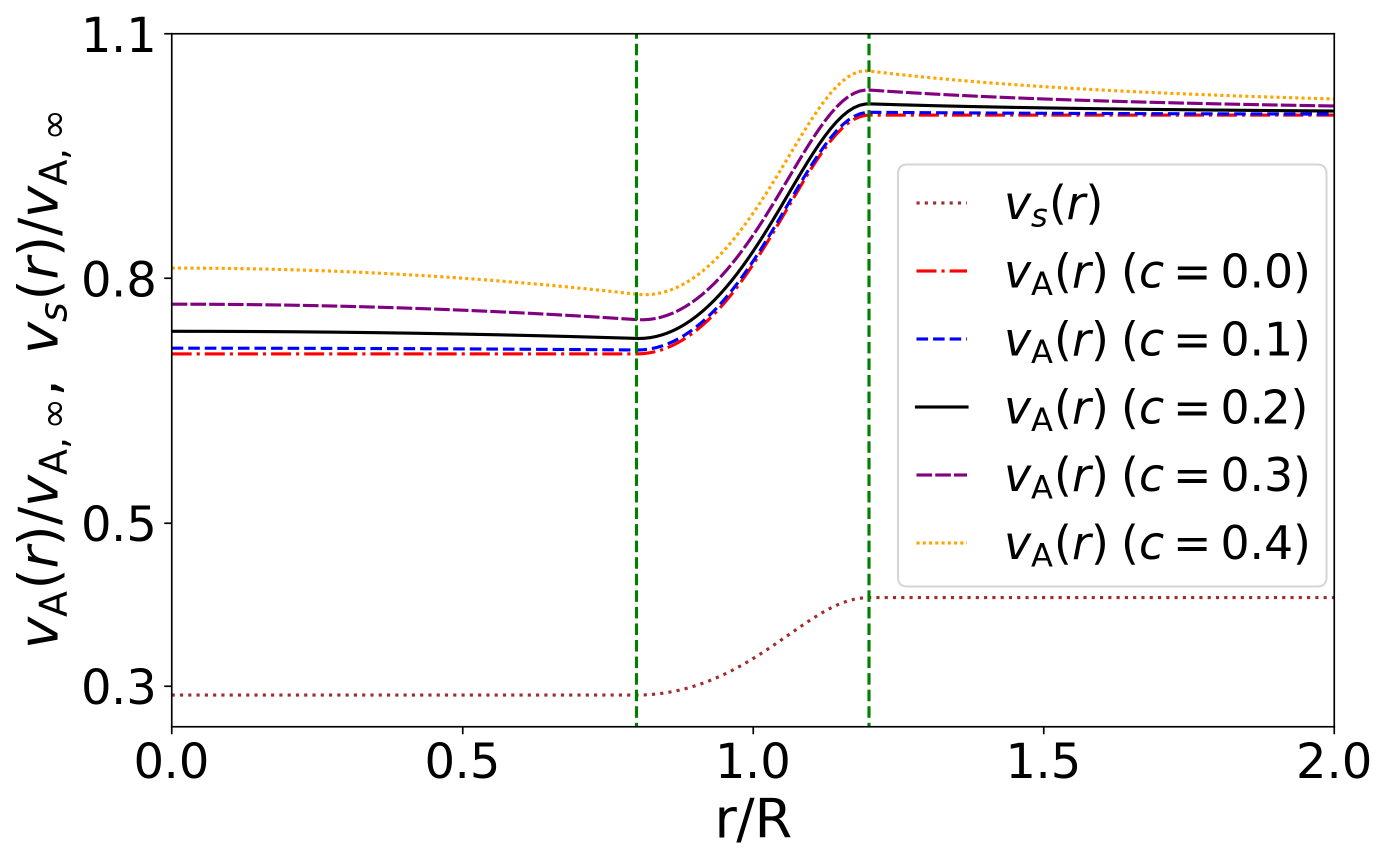}} 
\centering
\caption{Equilibrium radial  profiles of the Alfv\'{e}n speed, $ v_{\rm A}(r)$, and sound speed, $v_{s}(r) $, for the twist parameter ranging from $c =0$ to $c=0.4$. The sound speed is the same in all cases. The region between the  vertical dashed lines denotes the inhomogeneous region in density. The values are normalized with respect to the  Alfv\'{e}n speed at $r \to \infty$. We used $\rho_{\rm i}/\rho_{\rm e}=2$ and $ l/R=0.4 $.} 
\label{Fig_soundalfven}
\end{figure}

The equilibrium Alfv\'{e}n, $v_{A}(r)$, and sound, $ v_{s}(r) $,  speeds are
\begin{eqnarray}
v_{\rm A}(r) &=&\frac{B (r)}{\sqrt{\mu \rho_0(r)}}, \\
 v_{s}(r) &=&\sqrt{\frac{\gamma p_{0}}{\rho_{0} (r)}},
\end{eqnarray}
where  $\gamma$ is the adiabatic constant. Figure~\ref{Fig_soundalfven} shows the radial profiles of both speeds for a particular set of parameters. The  Alfv\'{e}n speed and sound speed both vary inside the transition region  because the density is nonuniform. However, when $c \neq 0,$ the Alfv\'{e}n speed is also position dependent in the inner core and in the external medium, where the density is uniform. This happens because the magnetic field is nonuniform everywhere when twist is present.

\begin{figure}[htbp!]
\resizebox{\hsize}{!}{\includegraphics{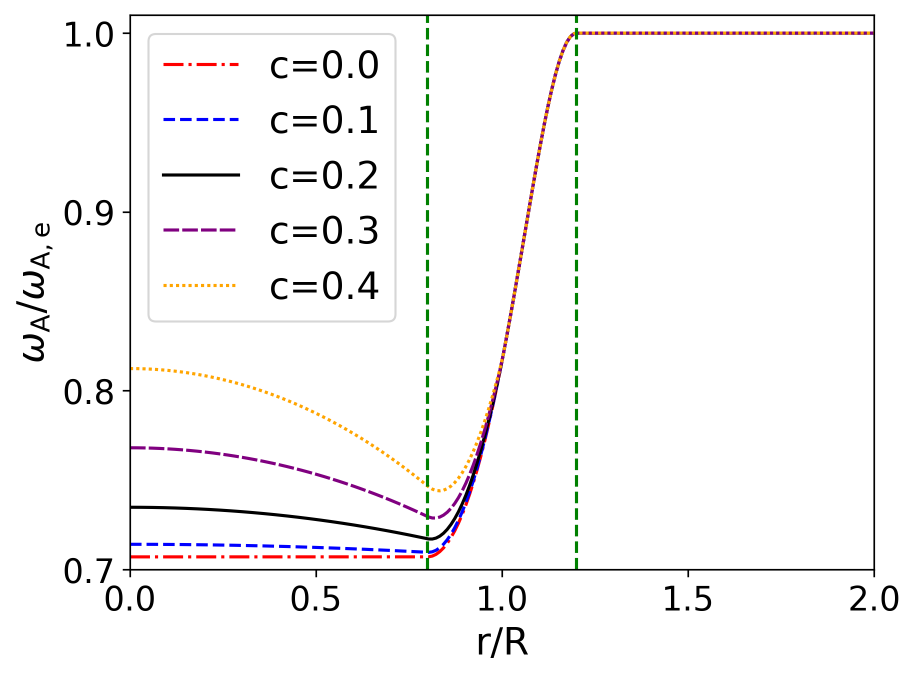}} 
\centering
\caption{Radial profile of the line-tied  Alfv\'{e}n frequency continuum for the twist parameter ranging from $c =0$ to $c=0.4$. The region between the vertical dashed green lines corresponds to the nonuniform region in density. The values are normalized to the external Alfv\'{e}n frequency. We used $\rho_{\rm i}/\rho_{\rm e}=2$ and $ l/R=0.4 $.}
\label{Fig_conti}
\end{figure}

According to \cite{Halberstadt93}, the line-tied Alfv\'{e}n frequency continuum is 
\begin{equation}
\omega_{\rm A}(r)=\frac{n\pi}{L}\frac{B_{z}(r)}{\sqrt{\mu_{0}\rho_{0}(r)}},
\label{conteq}
\end{equation}
where $n$ is the harmonic number. A continuous spectrum is present because of the radial variation of the $z$-component of the magnetic field, $B_z(r)$, and the density, $\rho_0(r)$. The density is only nonuniform in the transition region. However, $B_z(r)$ is also nonuniform in the inner core of the loop when twist is present. Because of twist, the Alfv\'{e}n continuum  also extends into the uniform core of the loop, as we show in Fig.~\ref{Fig_conti}.

Throughout the paper, length and density are normalized with respect to the radius of the flux tube, $ R $, and the external density, $ \rho_{e} $, respectively. Velocities are normalized with respect to the external Alfv\'{e}n speed at $ r \to \infty $, namely $v_{A,\infty} = v_{A}\left( r \to \infty \right)$. The normalized time is $ \overline{t} = t/\tau_{\rm A}$, with $\tau_{\rm A} =  R/v_{A,\infty} $. 

\subsection{Numerical code}

As in Paper~\citetalias{Diazsoler21b}, the numerical simulations were performed with the  PLUTO code \citep{Mignone07}. We  solved the 3D ideal MHD equations with a finite-volume shock-capturing spatial discretization on a structured mesh. The equations are as follows:
\begin{eqnarray}
\frac{\partial\rho}{\partial t} &=& -\nabla \cdot\left(\rho \mathbf{v}\right), \label{cont} \\
\rho \frac{\mathrm{D}\mathbf{v}}{\mathrm{D} t} &=&  -\nabla p + \frac{1}{\mu_0} \left(\nabla \times \mathbf{B}\right) \times \mathbf{B}, \label{mome} \\ 
\frac{\partial \mathbf{B}}{\partial t} &=& \nabla \times \left(\mathbf{v}\times \mathbf{B}\right), \label{indu} \\
\frac{\mathrm{D} p}{\mathrm{D} t} &=& \frac{\gamma p}{\rho}\frac{\mathrm{D} \rho}{\mathrm{D} t}. \label{ener} 
\end{eqnarray}
In Eqs. (\ref{cont})-(\ref{ener}),  $ \frac{\mathrm{D}}{\mathrm{D}t}  \equiv \frac{\partial}{\partial t}+ \mathbf{v} \cdot\nabla$ denotes the total derivative, $ \rho $ is the mass density, $ \mathbf{v} $ is the velocity,  $ \mathbf{B} $ is the magnetic field, $ p$ is the gas pressure, and  $\mu_{0}$ is the vacuum magnetic permeability. Gravity and nonideal terms are neglected.

The PLUTO code solves Eqs. (\ref{cont})-(\ref{ener}) in Cartesian coordinates. As in Paper \citetalias{Diazsoler21b}, the code uses a Roe-Riemann solver \citep{Roe81} to compute the numerical fluxes, a piecewise parabolic scheme for spatial reconstruction, and a second-order Runge-Kutta algorithm for temporal evolution. In order to maintain the solenoidal constraint, the generalized Lagrange multiplier (GLM) method \citep{Dedner02} is used. The background magnetic field  is not current-free because of magnetic twist, and therefore the background-field-splitting technique cannot be used in the present simulations, in contrast to Paper~\citetalias{Diazsoler21b}. For comparison purposes, the background-field-splitting technique is not used for the straight magnetic field case either. The code therefore advances the full magnetic field vector here.

We used the same base computational box as in Paper \citetalias{Diazsoler21b}, namely a numerical resolution of 100x100x100 cells distributed uniformly from $-3R$ to $3R$ in the $x$-  and $y$-directions, and from $-L/2$ to $L/2$ in the $z$-direction. The code implements the adaptive mesh refinement (AMR) strategy \citep{Mignone12} by which the cells split into two, and so the spatial resolution doubles if a certain criterion is satisfied. The refinement criterion is based on the second derivative error norm \citep{Lohner87} of a parameter closely related with the kinetic energy density because the goal of the AMR strategy in the present simulations is to conserve wave energy as much as possible. The use of AMR decreases the computational cost by keeping a low resolution where the dynamics is not relevant and generating a high resolution in the regions of interest. We included four levels of refinement,  so that the maximum effective resolution was 1600x1600x1600. For typical values of a coronal loop radius, the effective transverse resolution was $\sim$ 10 km.

Although the AMR scheme allowed us to deal with extremely fine scales,  the dynamics developed in the flux tube eventually cascaded the energy to scales below the maximum effective resolution. As in Paper \citetalias{Diazsoler21b}, to determine the simulation time at which this unphysical energy loss starts to happen, we monitored the kinetic, magnetic, and internal energies. In this way, we were able to determine the time at which numerical dissipation is beginning to play an important role. This time also coincides with the time when the total integrated vorticity saturates and decreases (see later). 

We set the same set of boundary conditions as in Paper \citetalias{Diazsoler21b}. We considered outflow conditions, that is, zero gradient for pressure, density, and the $x$- and $y$-components of the magnetic field at all boundaries. We also set the outflow condition for the $z$-component of the magnetic field, $ B_{z} $, at the lateral boundaries. To mimic the line-tying of the magnetic field lines at the solar photosphere, $ B_{z} $ was fixed to the equilibrium value through Eq. (\ref{bz}) at the top and bottom boundaries. The three velocity components vanish at all boundaries to avoid energy injection from outside the domain.

\subsection{Initial perturbation}

\begin{figure}[htbp!]
\resizebox{\hsize}{!}{\includegraphics{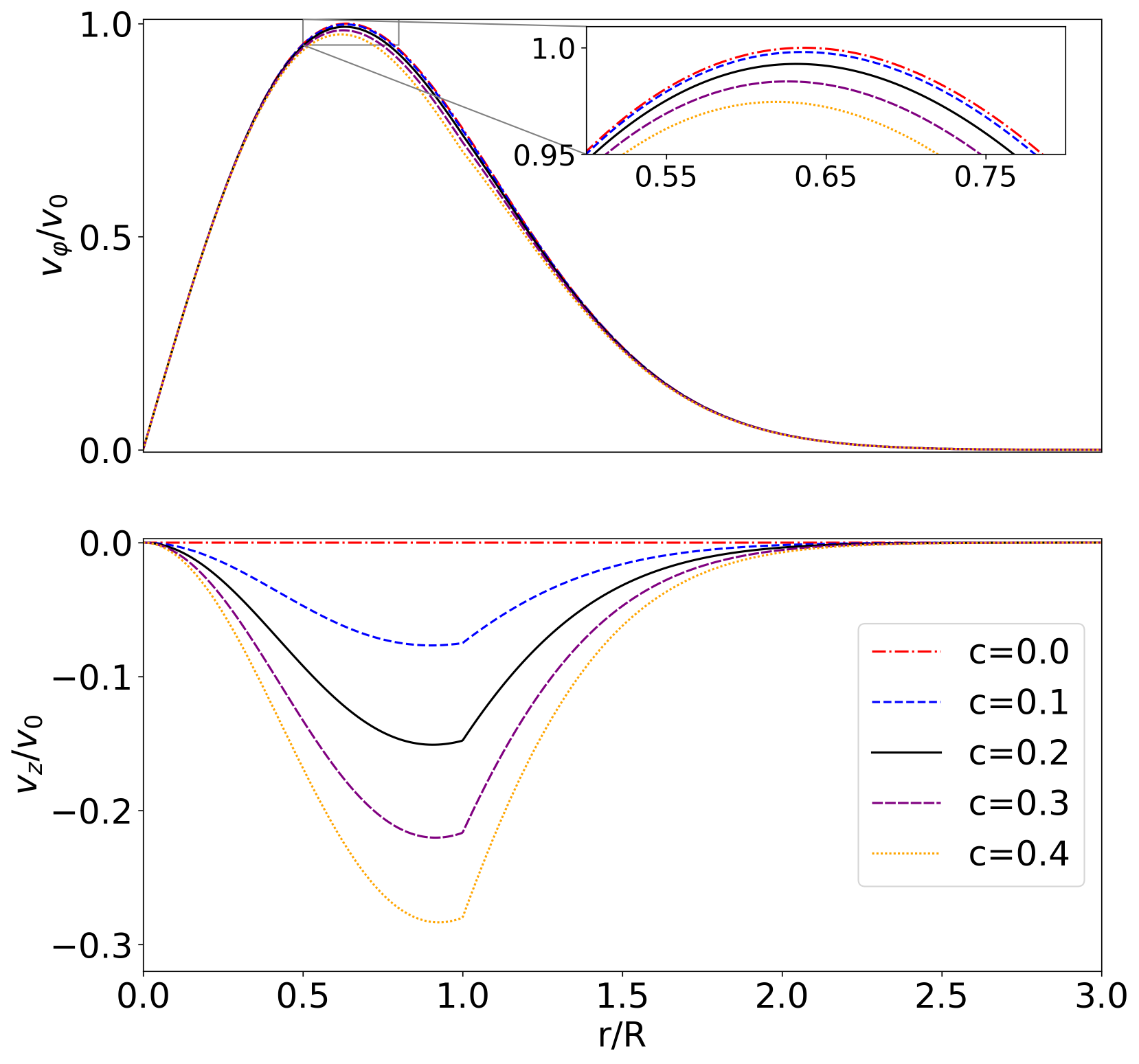}} 
\centering
\caption{Radial profiles of the azimuthal (top) and longitudinal (bottom) velocity components at $t=0$ for the twist parameter ranging from $c =0$ to $c=0.4$. The inset in the top panel shows a close-up view of the azimuthal velocity around its maximum. Both velocity components are normalized to the maximum amplitude  of the perpendicular velocity perturbation, $v_{0}$.}
\label{Fig_ampli}
\end{figure}

In the solar atmosphere, torsional Alfv\'{e}n waves can be driven by photospheric plasma motions \citep[see, e.g.,][]{Fedun11,Shelyag11,Shelyag12,Wedmeyer12,Mumford15,Srivastava17b}. The waves propagate  to the corona through the chrosmosphere, where reflection and dissipation heavily reduces the upward wave transmission \citep[see][]{Soler19}. When a broadband spectrum of  waves driven at the photosphere is transmitted into a closed coronal structure such as a coronal loop, \citet{Soler21} showed that most of the energy is deposited in the longitudinally fundamental mode. In view of this result and as in Paper~\citetalias{Diazsoler21b},  we excited the longitudinally fundamental mode of standing torsional Alfv\'{e}n
waves at $t=0$ . We did this by perturbing the components of velocity. To prevent the perturbation from also initially exciting slow MHD waves, we imposed an initial velocity field in the same manner as in \citet{Diazsoler21}, so that the velocity was purely perpendicular to the magnetic field lines, namely
\begin{eqnarray}
v_{r}(r,z) &=&0, \label{vvr} \\
v_{\varphi}(r,z) &=& v_{\perp} (r,z) \frac{B_{z}(r)}{B(r)} ,\label{vvphi}\\
v_{z}(r,z) &=&  -v_{\perp} (r,z) \frac{B_{\varphi}(r)}{B(r)},\label{vvz}
\end{eqnarray}
where $ v_{\perp}$ is the perpendicular velocity to the magnetic field lines.  We considered the same form as that used in Paper~\citetalias{Diazsoler21b}, namely
\begin{equation}
v_{\perp} (r,z) = v_0 A\left(r\right) \cos\left(\frac{\pi}{L} z\right),
\label{vperp}
\end{equation} 
where $v_0$ is the maximum velocity amplitude and  $A\left(r\right)$ contains the radial profile (see Paper~\citetalias{Diazsoler21b}). We set $v_0 = 0.1 v_{A,\infty} $. The radial profile of $ v_{\varphi} $ and $ v_{z} $ at $ z=0 $ is shown in Fig. \ref{Fig_ampli} for different values of the twist parameter, $c$.

We recover the same initial condition as in Paper~\citetalias{Diazsoler21b} for the untwisted case, $ c=0 $. When we increased the magnetic twist, the maximum of $v_{\varphi} $  slightly  decreased,  but its profile reamined mostly unaltered. The longitudinal component of velocity, $v_{z} $, increased significantly when the twist parameter grew. This is a direct effect of the relative variation of the azimuthal and longitudinal components of the magnetic field. 

Although the initial perturbation will mostly excite torsional Alfv\'{e}n waves, other types of MHD can also appear in the flux tube. The inclusion of a twisted magnetic field introduces linear coupling between torsional Alfv\'{e}n waves and fast magnetoacoustic sausage waves \citep{Sakurai91,Goossens92,Goossens11,Giagkiozis15}. The waves excited by the initial perturbation when $c\neq 0$ will develop mixed properties of both torsional Alfv\'{e}n waves and fast magnetoacoustic sausage waves. Nevertheless, because the twist is weak, the torsional Alfv\'{e}n wave character will be dominant. Later in the evolution, the ponderomotive force will additionally nonlinearly  drive  slow MHD waves \citep{hollweg1971,Rankin1994,Tikhonchuk1995,Terradas2004}.

\section{Revisiting the straight magnetic field case}
\label{Subsect_stra}

The main goal of this paper is to illustrate the effects of a twisted magnetic field in the nonlinear evolution of the standing torsional Alfv\'{e}n waves. In particular, we investigate its influence on the triggering of the KHi and the associated turbulence generation. However, before we investigate the twist, we revisit the  straight  magnetic field case of Paper \citetalias{Diazsoler21b} and perform an extended study.

We reran the simulation corresponding to the thin-layer case of  Paper~\citetalias{Diazsoler21b} but for longer times and deeper into the turbulent phase. The simulation was continued beyond the capacity of the AMR scheme in this way to correctly describe the smallest spatial scales that develop in the computational box. Although the dynamics of the smallest scales will then be affected by numerical dissipation, we can assume that the evolution at spatial scales larger than the maximum effective resolution will remain valid. Having this in mind, we can study how turbulence evolves globally at longer times despite the unreliable information at the smallest scales.

We refer to Paper~\citetalias{Diazsoler21b} for detailed explanations of the simulation evolution. We give a brief summary. Standing torsional Alfv\'{e}n waves excited in the flux tube develop the process of phase mixing \citep[see, e.g.,][]{HeyvaertPriest83,Nocera84,Moortel00,Smith07,Prokopyszyn2019,Ebrahimi2020}. In the transition region, where density is nonuniform, waves in adjacent magnetic surfaces oscillate with a slightly different frequency (see Eq.~(\ref{conteq})) and become out of phase as time increases. The effect of phase mixing is to generate an azimuthal velocity shear, which eventually triggers the KHi \cite[see, e.g.,][]{HeyvaertPriest83,Soler10,Zaqarashvili15,Barbulescu19}. Initially, the KHi develops locally in the transition region, which means that it does not globally perturb the flux tube. Nonetheless, the KHi evolves nonlinearly as time increases. The nonlinear evolution of the KHi breaks large eddies into smaller eddies. The instability naturally drives turbulence in the flux tube, which develops perpendicularly to the background magnetic field.  Turbulence extends beyond the boundaries of the initial nonuniform transition, and further accelerates the energy cascade toward small scales. Because of turbulence, the internal and external plasmas mix. Thus, the dynamics of torsional Alfv\'{e}n oscillations is similar to that of linearly polarized kink \citep[see, e.g.,][]{Terradas08,Antolin15,Magyar16,Karampelas19,Antolin19,Pascoe20,Martinez22} or circularly polarized kink oscillations \citep{Magyar22}. Nonetheless, the flux tube is not displaced laterally in the case of torsional oscillations, as occurs with the excitation of kink oscillations, and the resonant absorption process is absent for torsional Alfv\'{e}n waves \citep[see, e.g., ][]{Goossens11}.

\begin{figure*}[htbp!]
\resizebox{0.75\hsize}{!}{\includegraphics{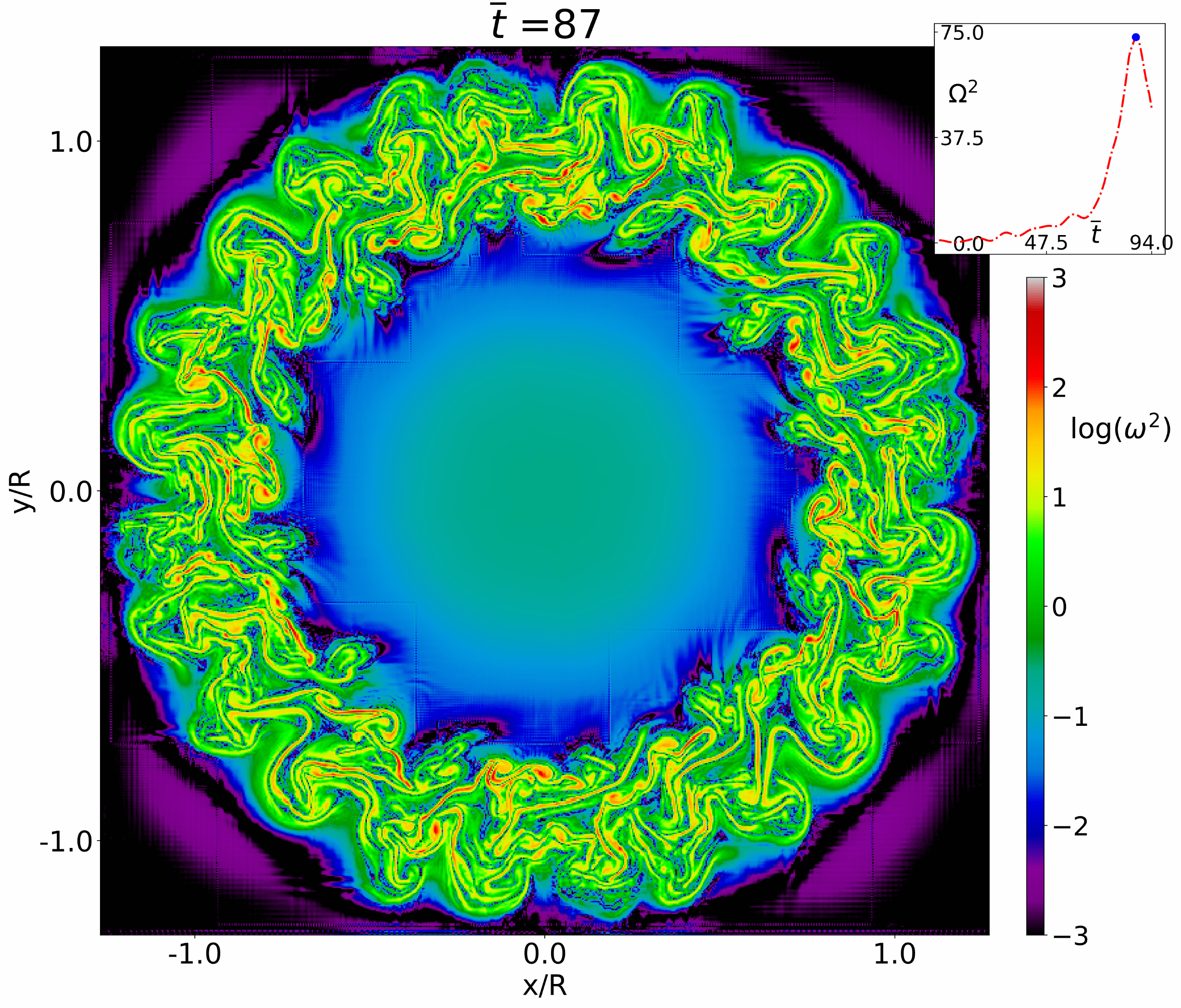}} 
\centering
\caption{Cross-sectional cut at $z=0 $ of the logarithm with base 10 of the vorticity squared, $\omega^{2}\left(\mathbf{r},t \right) $, in arbitrary units when $ \overline{t}=87 $.  The small panel to the right shows the temporal evolution of $\omega^{2}\left(\mathbf{r},t \right) $ integrated over the whole computational domain, $\Omega^{2}(t)$,  normalized with respect to the value at $\overline{t}=0$. The blue dot marks the position in the curve for the considered simulation time. The complete temporal evolution is available as a movie.} 
\label{Fig_vortynotwist}
\end{figure*}

 As shown in Paper~\citetalias{Diazsoler21b}, a  parameter that helps us understand the different phases of the evolution is the vorticity. We investigated the vorticity squared, $  \omega^{2} (\mathbf{r},t) = \left| \nabla \times \mathbf{v}(\mathbf{r},t)\right|^2$, and its value integrated over the whole computational domain, $ \Omega^{2} $, namely
\begin{equation}
\Omega^{2} \left(t\right)=\int \omega^{2}\left(\mathbf{r},t \right) \mathrm{d}^{3}\mathbf{r}.
\label{vortyintegrate}
\end{equation}
Figure \ref{Fig_vortynotwist} shows a cross-sectional cut at the tube center, $z=0 $, of $\omega^{2}\left(\mathbf{r},t \right) $ in  logarithmic scale. Only a subdomain of the complete numerical domain is shown, where the relevant dynamics occurs. The inset to the right shows the  integrated vorticity squared, $ \Omega^{2} \left(t\right)$, as a function of the computational time. The blue dot tracks the position on the curve of $ \Omega^{2} $ as time varies. The complete temporal evolution is available as a movie. 

In the small panel of Fig.~\ref{Fig_vortynotwist}, the temporal evolution of $ \Omega^{2} $ undergoes three phases (see Paper~\citetalias{Diazsoler21b}). In the first phase for $\bar{t} \lesssim 65$, $ \Omega^{2} $ oscillates and slightly increases owing to the linear development of phase mixing. In this phase, the spatial distribution of $\omega^{2}\left(\mathbf{r},t \right) $ is mostly in the form of concentric rings.  The concentric rings deform after the KHi is locally excited. In the second phase for $65 \lesssim \bar{t} \lesssim 87$, the KHi has grown enough to behave nonlinearly, and it induces turbulence. The KHi vortices break into increasingly smaller vortices. As a result, there is a dramatic increase in vorticity. The spatial distribution of $\omega^{2}\left(\mathbf{r},t \right) $ clearly reveals the formation of very fine scales owing to turbulence. Finally, in the third phase for $\bar{t} \gtrsim 87$, the AMR scheme can no longer fully describe the  smallest scales that appear in the evolution. Thus, vorticity decreases unphysically because of numerical dissipation. As a reference, the periods of the internal and external Alfv\'en oscillations are $20 \sqrt{2}$ and $20$ normalized time units, respectively.

The particular snapshot displayed  in the large panel of Fig.~\ref{Fig_vortynotwist} shows the spatial distribution of $\omega^{2}\left(\mathbf{r},t \right) $ at the time for which  $ \Omega^{2} $ is maximum. The loop boundary is clearly fully turbulent, and  extremely fine structures in vorticity have already been generated. Several authors have studied the evolution of vorticity in numerical simulations where the kink mode is excited \citep[see, e.g.,][]{Terradas08,Antolin15,Howson17res,Karampelas17,Karampelas18,Guo19,Howson2020,Howson2021} and some of their results are comparable with ours. Similar vorticity structures as those displayed in Fig.~\ref{Fig_vortynotwist} can also be seen in previous works of nonlinear kink oscillations \citep[see, e.g.,][]{Antolin14,Howson17a,Antolin19}. Nonetheless, the spatial distribution is  different due to the different azimuthal symmetry of the torsional and kink oscillations.

\begin{figure}
\resizebox{\hsize}{!}{\includegraphics{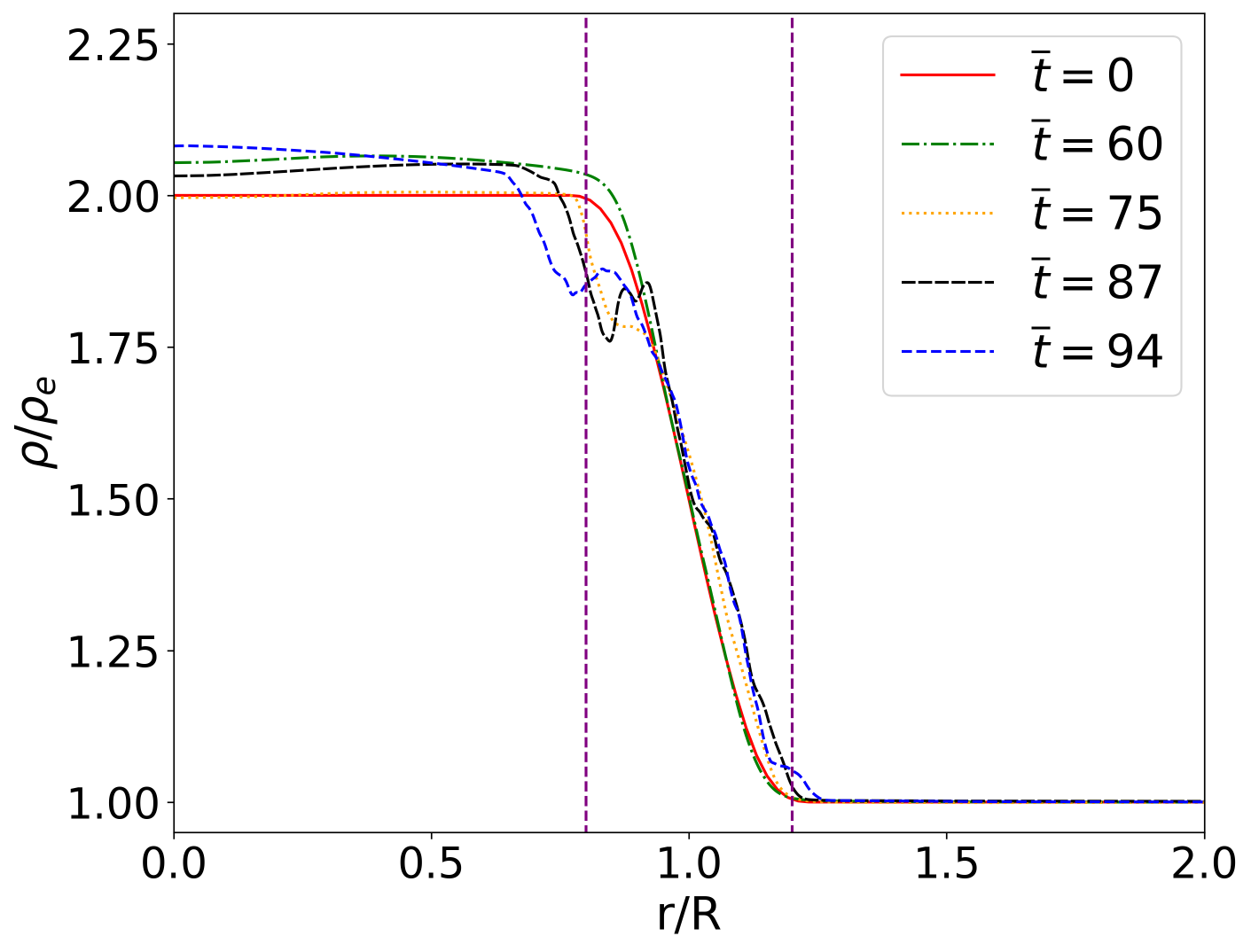}} 
\centering
\caption{Azimuthally averaged density at the tube center, $z=0$, for different simulation times. The vertical dashed purple lines show the limits of the nonuniform region at $\overline{t}=0$. Density is normalized to the external density.}
\label{Fig_averagedens}
\end{figure}

The turbulent evolution of the flux tube is clearly highlighted by the formation of extremely fine vorticity structures captured by the high resolution of the AMR scheme. A consequence of turbulence is the mixing of the internal and external plasmas, which heavily modifies  the  transverse density profile adopted initially. To illustrate this effect, Fig.~\ref{Fig_averagedens} displays an azimuthal average of the transverse density profile at $z=0$ for various times. Figure~\ref{Fig_averagedens} is similar to the density panels from Fig.~5 of \citet{Karampelas18} for  nonlinear kink waves. On average, we find that the width of the nonuniform region increases with time. For long times in our simulation, part of the initially uniform core becomes turbulent, and thus nonuniform. Consequently, at sufficiently long times beyond those simulated here, the whole flux tube should become turbulent. It is likely that for kink waves, which are global oscillations of the whole tube, the fully turbulent regime is reached faster than for torsional Alfv\'en waves, which are local waves in nature. This would require further analysis. We note that the density variations are
smoothed because of the azimuthal average. Along specific directions, we can find density variations that are far stronger than those displayed in Fig.~\ref{Fig_averagedens}. In addition, the full temporal evolution of the results displayed in Fig.~\ref{Fig_averagedens} also shows periodic density increases in the loop core caused by the ponderomotive force \citep[see, e.g.,][]{hollweg1971}.

\section{Exploring the effect of  magnetic twist}

Here we investigate the influence of  magnetic twist. In addition to the simulation without magnetic twist ($c=0$) discussed before, we performed four additional simulations including magnetic twist with $c=$~0.1, 0.2, 0.3, and 0.4. 

\begin{figure}[htb]
\includegraphics[width=\columnwidth]{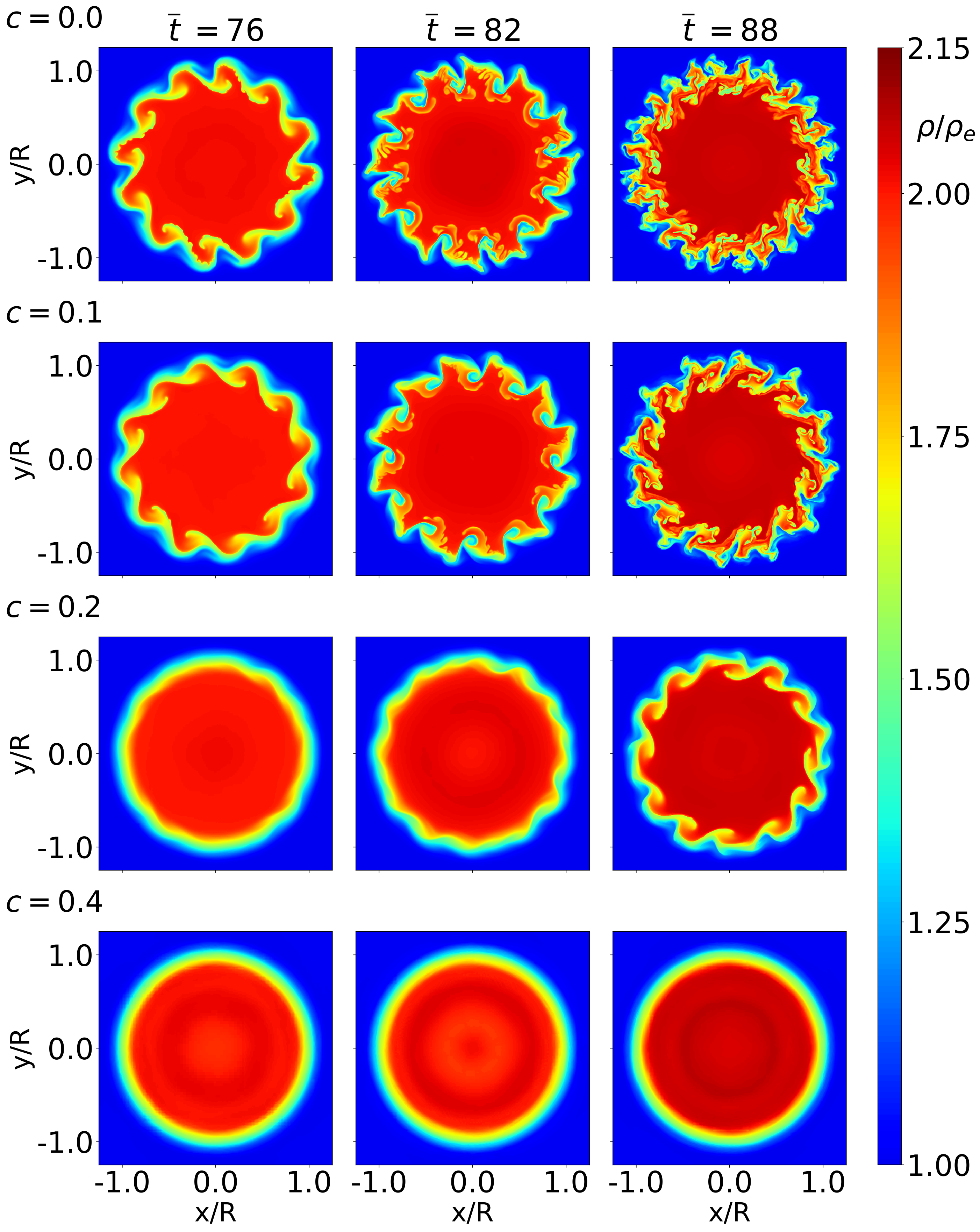} 
\centering
\caption{Cross-sectional cuts of density at the tube center, $z=0$, for different simulation times: $ \overline{t}=76 $ (left column), $ \overline{t}=82 $ (mid column), and  $ \overline{t}=88 $ (right column). The first row corresponds to the simulation without magnetic twist, $c=0$. In descending order, the remaining rows correspond to the simulations with magnetic twist parameters of $ c =$~0.1, 0.2, and 0.4. Density is normalized to the external density.} 
\label{Fig_Densitytwist}
\end{figure}

Before the onset  of the KHi, when the dynamics is governed by the development of phase mixing, the results of all simulations are quite similar, regardless of the strength of magnetic twist. The reason is that the shape of the Alfv\'en frequency continuum in the nonuniform layer is very similar in all cases (see Fig.~\ref{Fig_conti}). Therefore, we focus on later times. In Fig. \ref{Fig_Densitytwist} we show cross-sectional cuts of the density at the tube center, $z=0 $,  at three different simulation times corresponding to already advanced stages of the simulations. The results for $c=0$ are shown in the first row. The twist parameter increases from $c=0.1$ to $c=0.4$  from top to bottom in the remaining rows. The results for $c=0.3$ are not shown as they are visually indistinguishable from those for $c=0.4$. Figure~\ref{Fig_Densitytwist} suggests that the inclusion of the magnetic twist in our model delays the onset  of KHi. For the same time, the KHi vortices appear to be progressively less developed when the twist increases. 

\begin{figure}[htbp!]
\includegraphics[width=0.6\columnwidth]{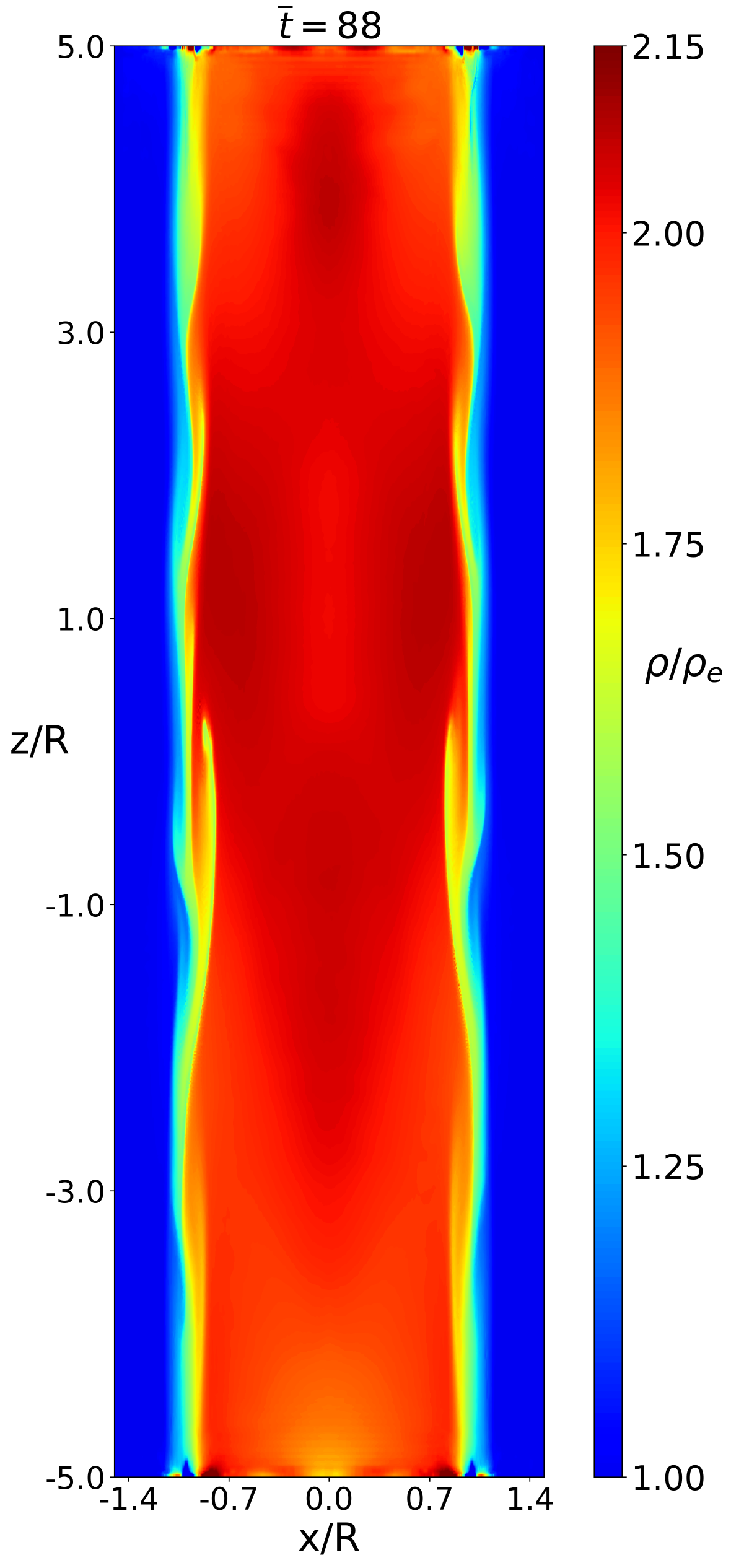}
\centering
\caption{Longitudinal density cut at $y=0$ for the simulation with magnetic twist parameter, $c = 0.2$ when $ \overline{t}=88 $, i.e., the same time as in the right column of Fig.~\ref{Fig_Densitytwist}.  Density is normalized to the external density.}
\label{Fig_rholong}
\end{figure}

On the other hand, when magnetic twist is included, the KHi vortices are not strictly perpendicular to the flux tube axis and can also develop in the $z$-direction \citep{Terradas18}. To illustrate this fact, we show in Fig.~\ref{Fig_rholong}  a longitudinal density cut of the simulation with $c=0.2$. This cut was made for the same time as in the third column of Fig.~\ref{Fig_Densitytwist}.

\subsection{Delay or suppression of the KHi}

We aim to quantify how magnetic twist postpone the KHi onset time.  Theoretically, a component of the magnetic field along the direction of the shear flow has a stabilizing effect on the KHi \citep[see, e.g.,][]{Chandrasekhar61}. Significant effort has been made to understand this problem for the KHi driven by transverse MHD waves. However, obtaining the onset time of the KHi in a cylindrical tube with time-varying flows under the presence of a twisted magnetic field remains a challenge analytically. \citet{Browning84} analytically obtained an expression for the onset time of the KHi for time-varying flows in Cartesian geometry and with a straight field. \citet{Soler10} considered the KHi at the boundary of a cylindrical tube, but with constant azimuthal flows. \citet{Soler10} studied the effect of magnetic twist by considering a Cartesian analog with an inclined magnetic field and found that twist can decrease the growth rates and even suppress the instability. \citet{Zaqarashvili15} studied the instability of a cylindrical, twisted, and rotating jet. They found that jets are unstable to the KHi only when the kinetic rotation energy is higher than the magnetic energy of the twist. \citet{Hillier2019MNRAS} investigated the linear stability of a discontinuous oscillatory shear flow  in Cartesian geometry in the presence of a uniform magnetic field. They showed that parametric instabilities can also grow in addition to the KHi. \citet{Barbulescu19} also considered a Cartesian model of time-varying flows, but including an inclined magnetic field on one side of the interface to mimic the effect of twist, as was previously done by \citet{Soler10} for constant flows. \citet{Barbulescu19} concluded that the magnetic shear may reduce the instability growth rate, but it cannot completely stabilize the interface, in contrast to the constant flow case of \citet{Soler10}.

To understand the effect of twist on the triggering of the KHi, we followed the method introduced by \citet{Terradas18}. Their method is based on studying the excitation of different azimuthal wave numbers. This approach was also used in \citet{Antolin19} and in  Paper~\citetalias{Diazsoler21b}. We proceeded as follows. In a cross-sectional cut at the tube center, $z=0$, we calculated the velocity component perpendicular to the magnetic field lines, $v_\perp$, in the middle of the transition region, $ r=R $, as a function of the azimuthal angle, $\varphi$, from $\varphi = 0$ to $\varphi =  2\pi$. After this, we computed the discrete Fourier transform to the data using the fast Fourier transform (FFT) algorithm with the Scipy module \citep{Virtanen20}. Following the notation of \citet{Terradas18}, the discrete Fourier transform is
\begin{equation}
G\left(p\right)=\sum^{N-1}_{k=0} g\left(k\right)\exp{\left(\frac{-2\pi i p k}{N}\right)},
\label{Gdep}    
\end{equation}
where $ N $ is the number of sample points, $ g\left(k\right) $ is the angular sampling of $v_\perp$, and $ p $ is an integer that plays the role of the azimuthal wavenumber. Unlike in Paper \citetalias{Diazsoler21b}, we considered positive and negative values of $p$, namely $ p=0,\pm 1, \pm 2, ... , \pm (N-1) $, since  twist breaks the degeneracy in the sign of the azimuthal wavenumber.   $p=0$ is the torsional plus the sausage mode. $ p=\pm 1 $ are the kink modes, and $ |p| \geq 2 $ are the fluting modes \citep[see, e.g.,][]{roberts19}. In the presence of magnetic twist, the torsional mode and the fast sausage mode are coupled linearly \citep[see, e.g.,][]{Goossens11}. However, the contribution of the fast sausage mode is weak because the magnetic twist considered here is relatively weak, even for $ c=0.4 $. Thus, for $p=0$, the torsional mode dominates the fast sausage mode.

\begin{figure}
\includegraphics[width=\columnwidth]{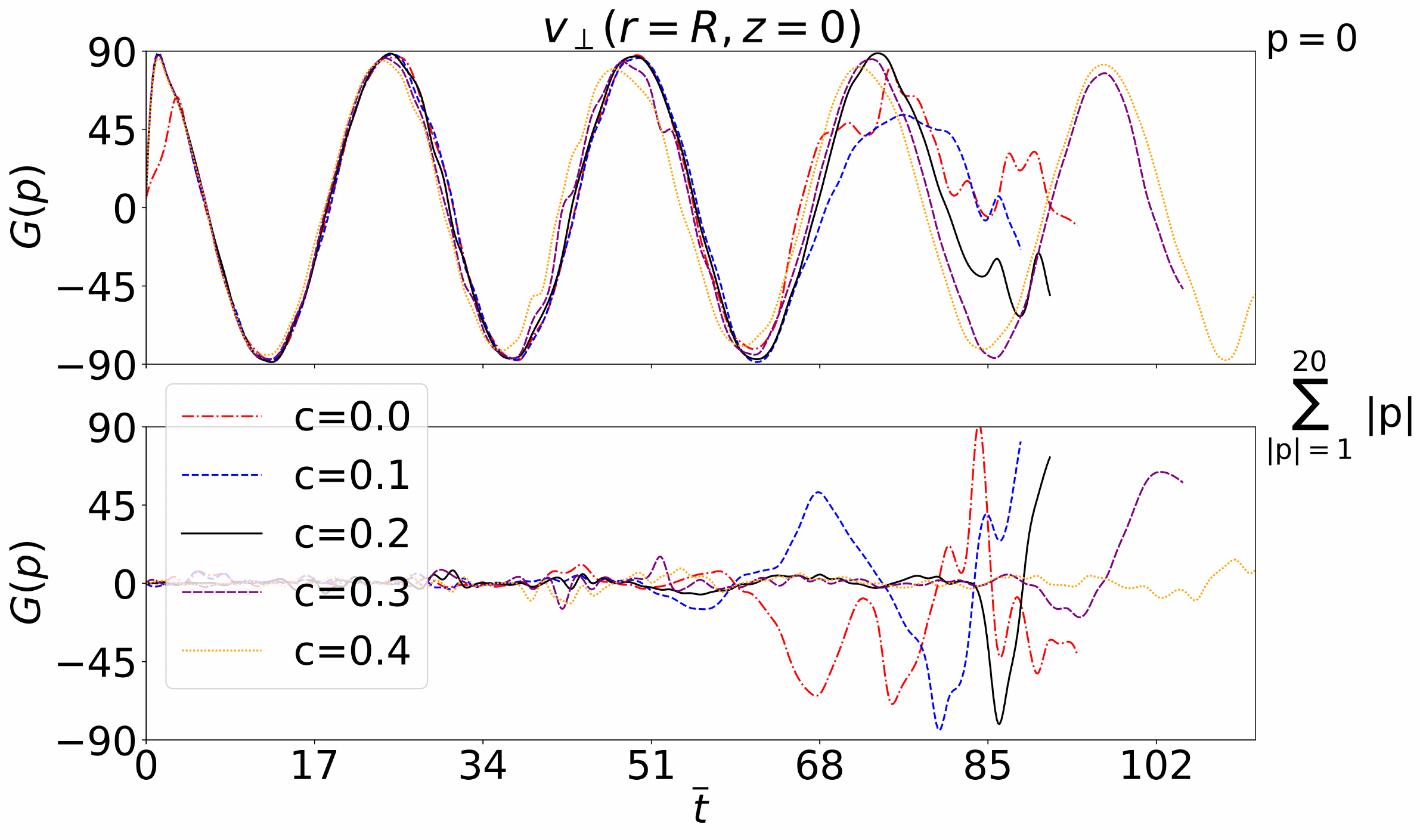}
\centering
\caption{Temporal evolution of the Fourier coefficients at the tube center, $z=0,$ and, in the middle of the inhomogeneous radial layer, $r=R$. \textit{Top}: The Fourier coefficient associated with the torsional mode, $ p=0 $. \textit{Bottom}: Same as the top panel, but for the sum of the first 20 positive and the first 20 negative Fourier coefficients excluding $p=0$. In both panels, the twist parameter ranges from $c=0$ to $c=0.4$. Arbitrary units are used.}
\label{Fig_Modos}
\end{figure}

Figure~\ref{Fig_Modos} shows the results of the azimuthal Fourier analysis. To better visualize  the results, we plot the temporal evolution of the $ p=0 $ mode alone in the top panel. In the bottom panel, we plot the temporal evolution of the sum of the first 20 positive modes and the first 20 negative modes, excluding the $p=0$ mode. We checked that adding more modes to the sum does not affect the results.

As expected, we obtain that the $p=0$ mode is dominant during the linear evolution of phase mixing regardless of the magnetic twist. The dominance of the $p=0$ mode lasts until the onset  of KHi, which excites higher azimuthal wave numbers. In this moment, the sum of the $p\neq 0$ Fourier coefficients begins to display significant variations. These results are consistent with those of \citet{Terradas18} and \citet{Antolin19} for the standing kink mode and also agree with those of Paper~\citetalias{Diazsoler21b}. During the development of KHi and the subsequent turbulence, the initially low amplitudes of the sum of the $p\neq 0$ Fourier coefficients become comparable with or even larger than the amplitude of the $p=0$ mode. Moreover, we find that the $p=0$ mode loses its periodicity.

Similarly to the results of Paper~\citetalias{Diazsoler21b}, regardless the presence or absence of magnetic twist, we verified that the excitation of high-order azimuthal wave numbers initially occurs in the nonuniform transition region. This analysis is not shown here for simplicity. We also verified the persistence of the $p=0$ mode in the core where the KHi cannot develop. However, when the turbulence expands into the core, high-order azimuthal wave numbers are also excited in the core.  This is consistent with the results from Fig. \ref{Fig_averagedens} for the untwisted case.  

Therefore, the azimuthal Fourier analysis confirms what was already visualized in Fig.~\ref{Fig_Densitytwist}: The inclusion of the magnetic twist  delays the development of the KHi. If the magnetic twist is sufficiently weak, the KHi is delayed, but the dynamics is similar to that in the absence of twist, although it evolves at a slower pace. In a sense, this conclusion is not completely new, although this is the first time that the role of twist  in the nonlinear evolution of standing torsional Alfv\'{e}n waves has been investigated. Similar results can be found in simulations of nonlinear kink waves in twisted flux tubes \citep[see, e.g.,][]{Howson17a,Terradas18}. For the different twist profiles considered by \citet{Howson17a}, the authors reported that the KHi vortices grow less as the magnetic twist is increased. \citet{Howson17a} also reported that the location of the maximum twist may affect at the development of KHi vortices. \citet{Terradas18} also studied nonlinear kink  waves considering the same twist profile as we used here. Similarly to \citet{Howson17a}, \citet{Terradas18} reported that the KHi vortices grow less when the magnetic twist increases.

When the  magnetic twist is strong enough, however, the physical scenario can be different. The simulations with strong magnetic twist suggest that the KHi can be suppressed altogether even if the phase-mixing-generated shear flows should remain unstable according to linear theory \citep[see][]{Barbulescu19}. Although the KHi itself can linearly be triggered in a local fashion, its perturbations are not allowed to fully grow into the nonlinear regime when the tension force provided by the twist is strong enough \citep{Diazsoler21}. The important role of magnetic tension in  nonlinearly preventing the growth of the KHi vortices has been discussed by \citet{Galinsky94}, \citet{Ryu2000}, and \citet{Hillier19}, for instance. In our model, the nonlinear suppression of the KHi seems to occur for $c\gtrsim 0.3$. 

Because the KHi suppression for strong twist is a nonlinear phenomenon, we cannot exclude that the initial amplitude of the excited torsional wave may affect the subsequent stability. For sufficiently large amplitudes, the KHi is expected to break the stabilizing effect of the magnetic tension even if the twist is strong. We therefore speculate that the ability of the KHi to grow might depend on the relation between the kinetic energy of the shear flow that drives the KHi and the magnetic energy of the twisted magnetic field, in a similar fashion as in the model by \citet{Zaqarashvili15}. The effect of the initial amplitude is not investigated here and is left for forthcoming works.

\subsection{Turbulence and vorticity generation}
\label{Sub_vortic_twist}

The delay or suppression of the KHi has important consequences in the dynamics of the flux tube.  In our model, the generation of turbulence, which results from the KHi evolution, can just be delayed or be absent, depending on the strength of magnetic twist. In addition, magnetic twist may affect the subsequent evolution of turbulence. To explore this, we resort again to the vorticity to study how magnetic twist affects the evolution of turbulence.

\begin{figure*}[htbp!]
\includegraphics[width=1.7\columnwidth]{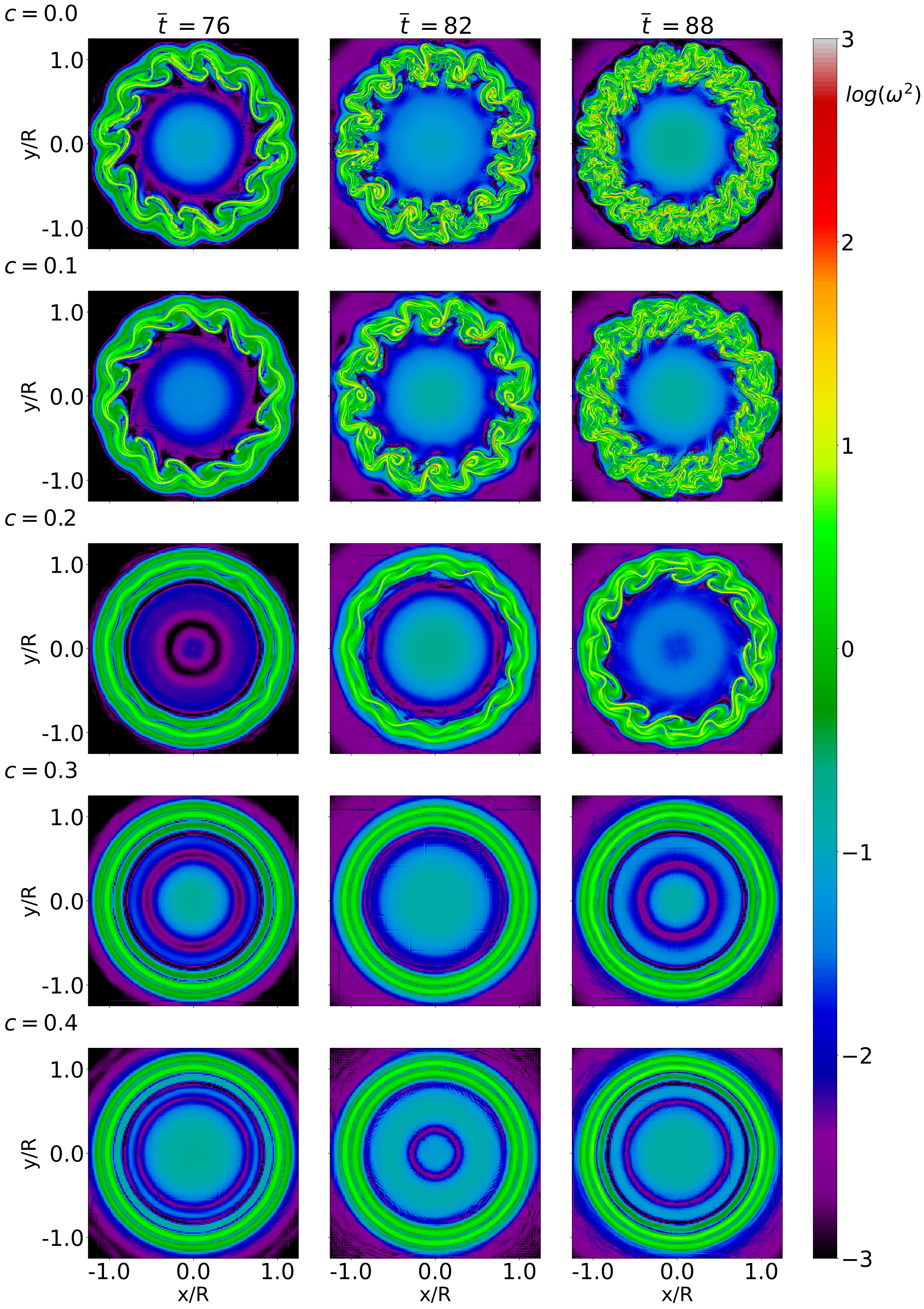}
\centering
\caption{Cross-sectional cuts at $z=0$ of the logarithm with base 10 of the vorticity squared, $\omega^{2}\left(\mathbf{r},t \right) $, for different simulation times: $ \overline{t}=76 $ (left column), $ \overline{t}=82 $ (middle column), and  $ \overline{t}=88 $ (right column). The first row corresponds to the simulation without magnetic twist, $c=0$. In descending order, the remaining rows correspond to the simulations with a magnetic twist parameter of $ c =$~0.1, 0.2, 0,3, and 0.4. Arbitrary units are used.}
\label{Fig_Vorty_local_twist}
\end{figure*}

\begin{figure}[htbp!]
\resizebox{\hsize}{!}{\includegraphics{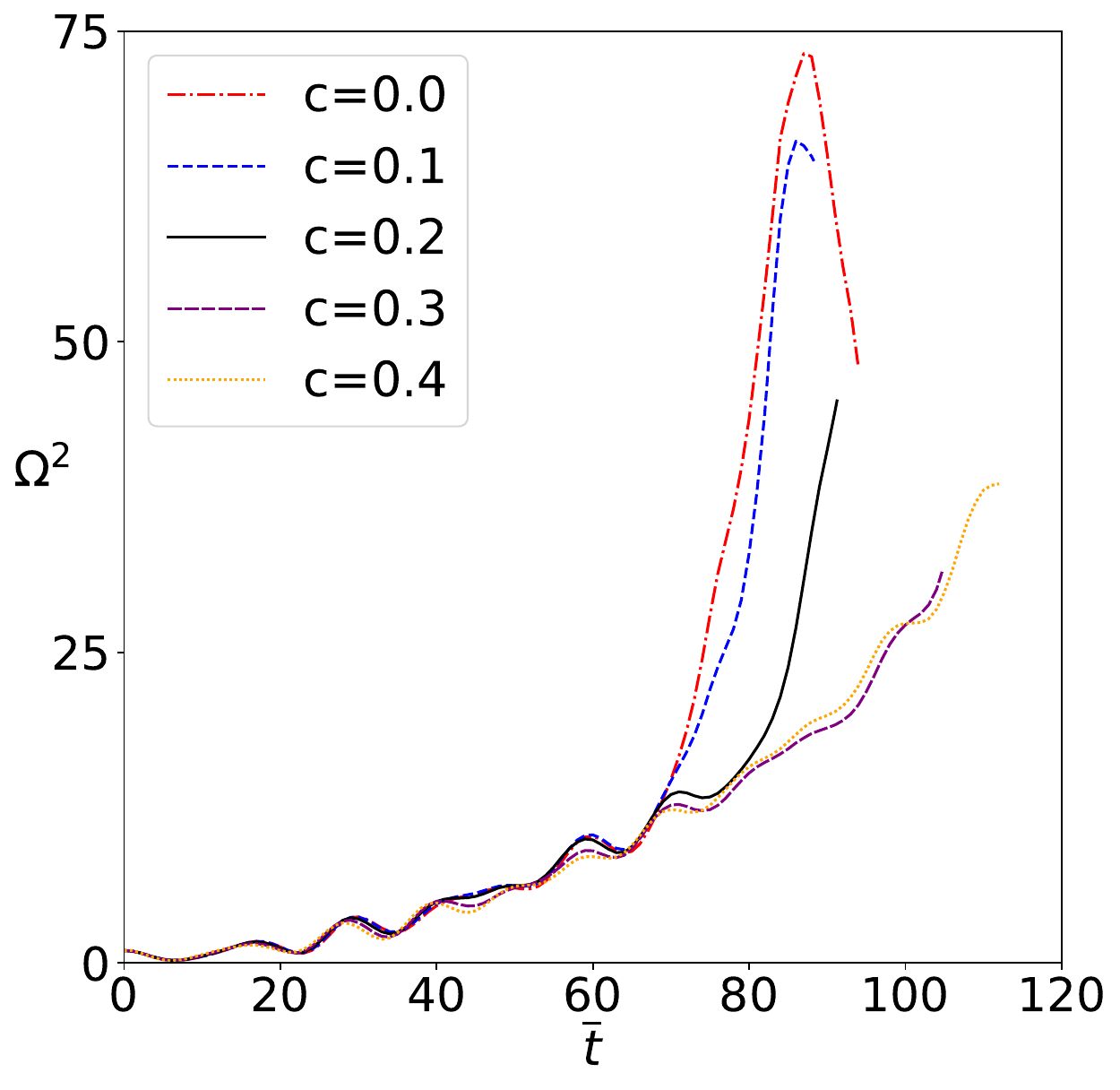}} 
\centering
\caption{Temporal evolution of the vorticity squared integrated in the whole computational domain, $ \Omega^{2} $, for the twist parameter ranging from $c =0$ to $c=0.4$. The curve for $c=0$ (dotted-dashed red line) is the same as in the small panel of Figure~\ref{Fig_vortynotwist}. All curves are normalized to their  values at $t=0$.}
\label{Fig_Vorty_twist}
\end{figure}

Figure \ref{Fig_Vorty_local_twist} shows cross-sectional cuts of $\omega^{2}\left(\mathbf{r},t \right) $ at the tube center, $z=0 $, for the  same three times as in Fig.~\ref{Fig_Densitytwist}. In each row, the strength of the magnetic twist is different and sorted in increasing order from  top to bottom. For comparison purposes, we also include the results from the simulation without magnetic twist in the first row. Unlike in Fig.~\ref{Fig_Densitytwist}, we find slightly different vorticity patterns for the cases with a magnetic twist of $c=0.3$ and $c=0.4$. This is highlighted below. Figure~\ref{Fig_Vorty_local_twist} shows that magnetic twist not only plays a role in delaying or suppressing the KHi, but also heavily affects the value and spatial distribution of vorticity when turbulence is present. Another plot that helps us in this discussion is Fig.~\ref{Fig_Vorty_twist}. There, we show the temporal evolution of the vorticity squared integrated in the whole computational domain, $ \Omega^{2} $, as a function of the computational time for all simulations. The curve for $c=0$ plotted in Fig.~\ref{Fig_Vorty_twist} is the same as was shown in the small panel of Fig.~\ref{Fig_vortynotwist}.

We start by focusing on comparing the simulations with $c=0$ (no twist) and $c=0.1$ (weakest magnetic twist considered). Figure~\ref{Fig_Vorty_twist} indicates that the onset  of the KHi occurs  practically at the same time in both simulations because the sharp increase in vorticity in the two cases is almost simultaneous. Thus,  the delaying effect of twist when $c=0.1$ is still not very relevant. However, slightly lower values of integrated vorticity are found when $c=0.1$ compared with those of $c=0$.
The first two rows in Fig.~\ref{Fig_Vorty_local_twist} show that the spatial distribution of vorticity when turbulence sets in is very similar in the two cases. Nevertheless, smaller vorticity structures appear when magnetic twist is absent. This effect is clearly visible in the two simulations when $ \overline{t}=88 $, a time for which turbulence is already well established. 

The effect of magnetic twist becomes more pronounced in the simulation with $c=0.2$.  Figure~\ref{Fig_Vorty_twist} shows that the KHi onset is significantly delayed with respect to the cases with $c=0$ and $c=0.1$ discussed before.  Figure~\ref{Fig_Vorty_local_twist} plainly shows that the dynamics  advances at a slower pace when $c=0.2$. For instance, the spatial distribution of $\omega^{2}\left(\mathbf{r},t \right) $ at $ \overline{t}=88 $ is  similar to that found at $ \overline{t}=76 $ in the simulation with $c=0.1$. The case with $c=0.2$ corresponds to an intermediate situation in the sense that twist already has an important impact in delaying the KHi onset  and in the appearance and evolution of turbulence. However, magnetic twist is still not strong enough to avoid the KHi growth and so prevent the generation of turbulence.

The simulations with stronger magnetic twist, $c=0.3$ and $c=0.4$, represent a completely different scenario in which turbulence is not driven because the KHi is not allowed to grow. Even for the longest considered simulation times, the shape of $\omega^{2}\left(\mathbf{r},t \right) $ in the two simulations is essentially in the form of  concentric rings, which is the typical structure caused by phase mixing. However, when $ \overline{t}=88, $  the shape of $\omega^{2}\left(\mathbf{r},t \right) $ in the case with $c=0.3$  reveals that some of the phase-mixing-driven concentric rings are slightly deformed in a wavy fashion. No such deformation is seen in the simulation with $c=0.4$ at that time. This deformation in the vorticity rings  when $c=0.3$ is caused by the very local triggering of the KHi within the nonuniform transition layer. This local inception of the KHi is not appreciable as discernible deformations in density. Nonetheless, vorticity, which is computed from the velocity field, is much more sensitive to the very initial appearance of the KHi.

We ran the simulation with $c=0.3$ up to longer times than those displayed in  Fig.~\ref{Fig_Vorty_local_twist} and confirmed that the KHi does not grow and no vortices form. As explained before, the reason is the nonlinear suppresion of the KHi by magnetic tension \citep[see, e.g.,][]{Galinsky94,Ryu2000,Hillier19,Diazsoler21}. Thus, turbulence is not driven. In the simulation with $c=0.4$, the local inception of the KHi is eventually observed at later times, but the perturbations are not allowed to grow  and turbulence is equally absent. However, Fig.~\ref{Fig_Vorty_twist} shows that vorticity continues to increase for both $c=0.3$ and $c=0.4$ at long times. The reason is that the linear phase mixing remains at work for long times and causes the increase in vorticity. Nevertheless, the increase in vorticity owing to phase mixing alone occurs at a much slower pace than when the KHi is present.

Again, our results about the effect of magnetic twist on the vorticity are comparable with those obtained from nonlinear kink wave simulations \citep{Howson17a,Terradas18}. \citet{Howson17a} showed cross-sectional cuts of the magnitude of vorticity at the loop apex for a twisted and an untwisted case at two different simulation times. In both cases, they reported large and small vortices due to KHi in their untwisted case, but their twisted case did not show small vortices. Thus, the magnetic twist suppresses the small vortices associated with KHi. \citet{Howson17a} also found that the increase in vorticity due to KHi occurs earlier when twist is absent.

\subsection{Kinetic energy cascade to small scales}

In Paper~\citetalias{Diazsoler21b} we showed that the onset of turbulence in the flux tube accelerates the cascade of energy from large to small spatial scales. This process is slower before the appearance of turbulence, when phase mixing is the only working mechanism. The delay or even suppression of the turbulent regime when magnetic twist is included should affect the rate at which small scales are generated.

We investigated how kinetic energy is transported to small scales depending on the strength of twist. For this purpose, we only considered the kinetic energy associated with motions perpendicular to the magnetic field lines, namely $ E =\frac{1}{2}\rho v^{2}_{\perp}$. We excluded the contribution from the other components of velocity because we are mainly interested in the Alfv\'enic part of the energy. The presence of  twist linearly couples  torsional Alfv\'{e}n waves and  fast magnetoacoustic sausage waves \citep[see, e.g.,][]{Sakurai91}. Fast magnetoacoustic sausage waves would predominantly perturb the radial velocity component. In turn, Alfv\'en waves and slow magnetoacoustic waves are nonlinearly coupled through the ponderomotive force \citep[see, e.g.,][]{hollweg1971}. Slow waves would mainly affect the longitudinal velocity component.

We  considered  a cross-sectional cut at the tube center, $z=0$, and averaged the Alfv\'{e}nic kinetic energy in the azimuthal direction using 512 radial cuts. Thus, the angular resolution is approximately $ \sim 0.7^{\circ}$. We denote the averaged energy by $ \overline{E}$, which is a function of $r$ and $t$. Then, we applied the continuous Fourier transform in the radial direction, discretized due to the limited numerical resolution. Following Paper~\citetalias{Diazsoler21b}, it can be defined as
\begin{align}
\overline{E}_F(k,t)\approx\frac{\Delta r}{\sqrt{2\pi}}\exp{\left(-i k r_{0}\right)} \sum^{N-1}_{m=0} \overline{E}(r,t)\exp{\left(-\frac{2\pi i m k}{N}\right)},
\label{fftcont}
\end{align}
where $N$ is the number of samples, $ k $ is the radial wave number, $ \Delta r $ is the spatial resolution, and $ r_{0} $ is the upper limit of the radial domain. The summatory in Eq.~(\ref{fftcont}) is exactly the 1D discrete Fourier transform, which we computed using the fft module of Scipy \citep{Virtanen20} based on the FFT algorithm \citep{Cooley65}. Finally, we calculated the modulus of $ \overline{E}_F(k,t) $ and normalized it with respect to the maximum value in the spectrum at each time.

\begin{figure}[htb!]
\resizebox{\hsize}{!}{\includegraphics{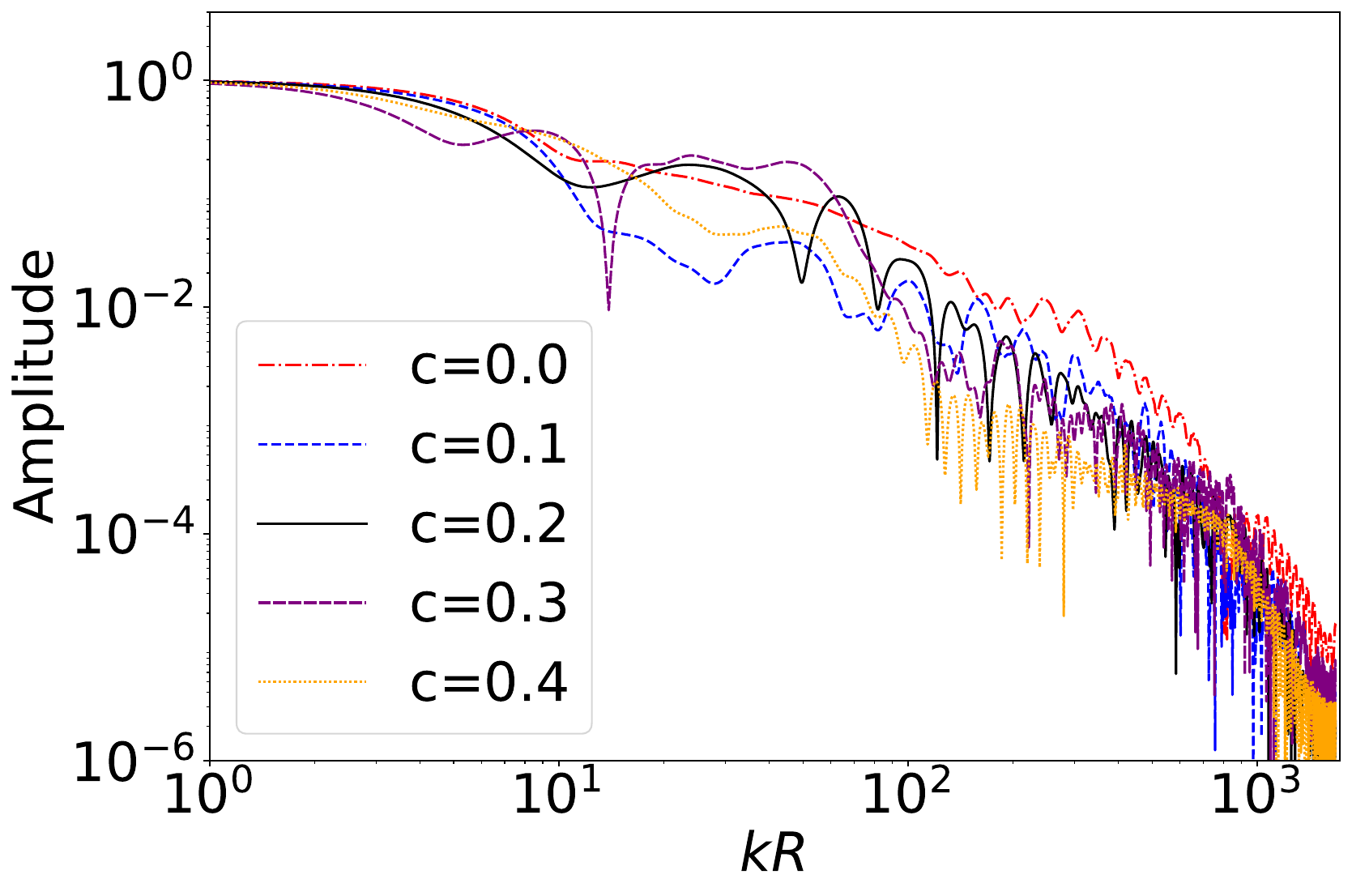}} 
\centering
\caption{Azimuthally averaged amplitude spectrum of the Alfv\'enic (perpendicular to the magnetic field) part of the kinetic energy for the twist parameter ranging from $c =0$ to $c=0.4$. We considered the last common  time for which vorticity increases in all simulations, $\bar{t}=87$. Arbitrary units are used.}
\label{Fig_Scales}
\end{figure}

Figure~\ref{Fig_Scales} displays the results of the Fourier transform for all simulations at the last common simulation time for which the vorticity still increases in all simulations, $ \overline{t}=87 $. The reason for choosing this time is that for longer times, the simulations with $c=0$ and $c=0.1$ are already significantly affected by numerical dissipation (see the decrease in integrated vorticity in Fig.~\ref{Fig_Vorty_twist}), which may  also affect the results of the Fourier transform at small scales.

At large scales, $kR < 10^0 $, the amplitude spectrum is independent of the wave number. For intermediate scales, $ 10^0 < kR < 10^2 $, we do not find that the values of the amplitude spectrum are ordered in any particular way regarding the twist parameter, $c$. The temporal evolution of the amplitude spectrum (not shown here) shows oscillations by which the various curves with different $c$ exchange their relative positions depending on the considered time. The reason for this behavior is unclear to us. Understanding of the behavior of the amplitude spectrum in these intermediate scales requires additional research.

We are more interested in the spectrum at small scales, $ 10^2 < kR < 10^3 $, where the amplitude spectrum decreases as the wavenumber increases. In the absence of  twist, the values of the amplitude spectrum are higher than those obtained when  twist is considered. In addition, we find a decreasing trend with respect to the twist parameter, $c$, so that the stronger the magnetic twist, the lower the values of the amplitude spectrum. These results again confirm the relevant role of magnetic twist in suppressing the generation of small scales and in the rate at which Alfv\'enic energy is deposited into those small scales. We intentionally avoided $ kR >  10^3 $ in our analysis owing to the possible role of numerical dissipation at the smallest scales.

\section{Concluding remarks}
\label{Sect_con}

We concluded the investigation started in Paper~\citetalias{Diazsoler21b} into the ability of torsional Alfv\'en waves to drive turbulence in coronal loops. Because of plasma and/or magnetic nonuniformity, standing torsional Alfv\'{e}n waves undergo phase mixing, generating shear flows perpendicular to the magnetic field direction in adjacent radial positions as time increases.  Eventually, the shear flows trigger the KHi, as \citet{HeyvaertPriest83} and \citet{Browning84} first predicted.

In the absence of magnetic twist (the case studied in Paper~\citetalias{Diazsoler21b}), the KHi can grow nonlinearly without opposition. The nonlinear evolution of the KHi naturally induces turbulence. First, large KHi eddies are formed, which later break into smaller eddies in a cascade-like dynamics. Mixing of plasma occurs.  An important increase in vorticity is found, and the generation of small scales speeds up compared with the initial phase, which is dominated by phase mixing alone. Turbulence evolves perpendicularly to the magnetic field, so that for a straight magnetic field, turbulence is pseudo-2D.

We have explored the role of magnetic twist. While the linear evolution of phase mixing is similar in the presence or absence of twist, the dynamics of the subsequent KHi growth and turbulence generation strongly depends on the strength of the twist.

If magnetic twist is sufficiently weak, the onset  and growth of the KHi is just delayed, but the dynamics is similar to that in the case of a straight magnetic field. In this regime, the stronger the magnetic twist, the longer the delay.  Although the vorticity still increases dramatically during the development of  turbulence, the increase is smaller than in absence of magnetic twist. Small scales are still quickly generated by turbulence, although at a slower pace than when twist is absent. Turbulence still evolves perpendicularly to the magnetic field lines, but because the field is twisted, turbulence is no longer confined to perpendicular planes to the tube axis. 

Conversely, when the magnetic twist is strong enough, the scenario is completely different. If the strength of  magnetic twist surpasses a critical value or a critical twist angle, no KHi vortices can grow. The KHi itself is still locally excited by the phase-mixing-generated shear flows, but the tension of the twisted magnetic field now prevents the nonlinear development of the instability \citep[see, e.g.,][]{Galinsky94,Ryu2000,Hillier19,Diazsoler21}.  As a consequence, there is no enhancement of vorticity, nor is turbulence driven.  Thus, the generation of small scales is not accelerated and continues to evolve slowly at the rate dictated by phase mixing. 

Although the effect of magnetic twist may oppose the process, the nonlinear evolution of torsional Alfv\'en waves remains a viable mechanism to induce turbulence as long as coronal loops are weakly twisted. Other effects should be explored in the future. For instance, the tension of the background field in curved coronal loops may also affect the triggering and evolution of the KHi and associated turbulence. It might be also interesting to investigate turbulence generation in multithreaded loops \citep[see, e.g.,][]{Ofman09}.

\begin{acknowledgement}
This publication is part of the R+D+i project PID2020-112791GB-I00, financed by MCIN/AEI/10.13039/501100011033. S.D. acknowledges the financial support from MCIN/AEI/10.13039/501100011033 and European Social Funds for the predoctoral FPI fellowship PRE2018-084223. We also thank the UIB for the use of the Foner cluster. For the simulation data analysis we have used VisIT \citep{VisIt} and Python 3.6. In particular, we have used Matplotlib \citep{Hunter07}, Scipy \citep{Virtanen20}, and Numpy \citep{Harris20}. We are thankful to B. Vaidya and his contributors for the tool pyPLUTO.
\end{acknowledgement}

\bibliographystyle{aa}
\bibliography{thebiblio} 
\end{document}